# KRC: KnowInG crowdsourcing platform supporting creativity and innovation

[1]Fernando Ferri, [2]Patrizia Grifoni, [3]Maria Chiara Caschera, [4]Arianna D'Ulizia, [5]Caterina Praticò

*[1,First Author] National Research Council - Institute of Research on Population and Social Policies (CNR-IRPPS), fernando.ferri@irpps.cnr.it*

*[2,Corresponcing Author] National Research Council - Institute of Research on Population and Social Policies (CNR-IRPPS), patrizia.grifoni@irpps.cnr.it*

*[3,4,5] National Research Council - Institute of Research on Population and Social Policies (CNR-IRPPS), mc.caschera@irpps.cnr.it, arianna.dulizia@irpps.cnr.it, caterina.pratico@libero.it*

## *Abstract*

*The deep financial and economic crisis, which still characterizes these years, requires searching for tools in order to enhance knowledge sharing, creativity and innovation. The Internet is one of these tools that represents a practically infinite source of resources. In this perspective, the KnowInG project, funded by the STC programme MED, is aimed at developing the KnowInG Resource Centre (KRC), a sociotechnical system that works as a multiplier of innovation. KRC was conceived as a crowdsourcing platform allowing people, universities, research centres, organizations and companies to be active actors of creative and innovation processes from a local to a transnational level.*

## 1. Introduction

The wide use of the Internet at a personal and an organizational level to deliver information, to share knowledge, to provide and to use services, is deeply changing the innovation of processes and products, as well as business processes both at regional and transnational level, supporting each participant in the global competition. The Internet, an almost infinite source of resources, has changed companies' knowledge management approach. In fact, they sometimes find convenient to widely involve people in sharing very specialized knowledge and in matching knowledge to the people who need it. In this perspective, traditional Knowledge Management Systems turn out to be insufficient to stimulate innovation and to make companies competitive at transnational level, as these systems are inside to the organization. This criticism is more relevant for Small and Medium Enterprises (SMEs) because of their restricted number of people in their staff and their connections with a single given territory, with the consequent difficulty to possess all the necessary competencies and knowledge at each time. Companies increasingly need to use external ideas, knowledge and technologies (resources) in order to push their activities and business. "Crowdsourcing" is a concept which is strongly related to the concept of external and shared resources and, as Estellés-Arolas et al. [1] underline, Internet Crowdsourcing particularly is identified as being a resource: "any type of Internet-based collaborative activity, such as co-creation or user innovation".

According to Zhao and Zhu [2], crowdsourcing is defined as being a "collective intelligence system" composed by three elements: an organization which directly takes advantage of the crowd's work; the crowd itself; and a platform which is able to connect the organization and the crowd, and to provide a host for the activity throughout its lifecycle. Crowdsourcing is frequently associated to the concept of Open innovation; according to Chesbrough et al. [3] "Open innovation is the use of purposive inflows and outflows of knowledge to accelerate internal innovation, and expand the markets for external use of innovation, respectively. This paradigm assumes that firms can and should use external ideas as well as internal ideas, and internal and external paths to market, as they look to advance their technology". In [4], crowdsourcing is "an emerging technique outside of its application by firms in product development cycles". Crowdsourcing platforms can be seen as open innovation tools allowing people, companies and organizations to improve and to increase the effectiveness of knowledge sharing and innovation. This can be performed by involving relevant external expertise and promoting cross-fertilization between different companies and organizations according to a trans-disciplinary approach. On the basis of these features, this paper presents an intelligent Web platform, named KnowInG Resource Centre (KRC), that has the purpose to engage the "crowd" in the co-creation and assimilation of resources, and to allow actors of innovation to identify problems and needs, matching them with the existing solutions and resources. This platform provides information and services to enterprises, research centres, universities, public and local administrations and people involved in creative and knowledge production processes in order to support innovation and creativity. This platform has been developed during the project "Knowledge Intelligence and Innovation for a sustainable Growth (KnowInG)", where crowdsourcing is a crucial concept.

KRC empowers the Web platform PLAKSS, "Platform of Knowledge and Services Sharing" which was designed and developed by the Social Informatics and Computing Unit from CNR-IRPPS of Italy.

The KnowInG project (with its activities and presence in some Mediterranean regions) has represented an important experience to promote the spreading of information and knowledge, and to stimulate social, cultural and economic innovation by encouraging the creation of new knowledge in an informed way. KnowInG was funded by the STC programme MED, along the Axe "Strengthening innovation capacities" and the Objective "Strengthening strategic cooperation between economic development actors and public authorities". The project allowed to share and experiment tools as well as to develop new innovation tools from regional to national and transnational levels. The KnowInG project provided services, actions and objects that allowed policies and actors supporting innovation.

This project did not only use and test tools that were already used in several regions as well as at European level to boost the growth of a knowledge economy; the effort was also aimed at creating, collecting and making available services, information and knowledge (e.g. resources) by gathering existing tools and new tools into the KRC. This system has allowed to express creativity and innovation throughout the entire project. In line with Dearden and Rizvi [5], KRC has been conceived in the perspective that "ICT for development is inescapably a socio-technical phenomenon". Moreover, Coakes and Coakes [6] said that "*Sociotechnical thinking is important to the design, development, implementation, and use of information technology systems in organizations. It addresses vital issues in combining the use of powerful information and communication technologies with effective and humanistic use of people… Its original emphasis was on organisational design and change management.*"

KRC is actually a sociotechnical system implemented as an intelligent Web platform that acts as a multiplier of innovation: 1) by aggregating and allowing to easily access and share services, information and knowledge already available in pre-existing tools according to an organized and unified approach; 2) by allowing each user to create her/his personal repository of services, information and knowledge, which can be shared with other KRC's users. The strength of the KRC relies on the fact that it is a hub of information, data, knowledge and services for the community of interest, and it enables people, organizations and companies to be active actors: they are both the problem creators and problem solvers.

KRC supports regional development in the following ways:

- by stimulating creative and innovation processes in the MED area,
- by involving people (crowd) and stakeholders on innovation in identifying problems and needs,
- by matching them with existing solutions and resources that can be involved in these solutions,
- by creating communities and networks that are the social structure on which the sociotechnical system is built.

The paper is structured as follows. The related researches on crowdsourcing and knowledge management systems, as elements of sociotechnical systems, will be analyzed in the next section. After that, a description of the KRC will be given. A discussion concludes the paper.

## 2. Crowdsourcing

In [7] a system can be defined crowdsourcing "if it enlists a crowd of humans to help solve a problem defined by the system owners" and if it focuses on the four fundamental questions: "How to recruit and retain users? What contributions can users make? How to combine user contributions to solve the target problem? How to evaluate users and their contributions?".

For the purpose of this paper, we will provide short descriptions of some crowdsourcing systems developed in the literature focusing on the following question "What contributions can users make?". This question focuses on functionalities of the systems that are offered to users, such as evaluation, sharing, networking, building artifacts, and executing tasks.

Crowdsourcing systems for evaluation allow users evaluating something, e.g. products, web pages, knowledge, users, using textual comments, numeric scores, or tags. Examples are Amazon, tagging Web pages at del.ici.ous.com and Google Co-op.

Crowdsourcing systems addressing sharing activities allow users to share structured knowledge, textual knowledge, products and services. Some of these systems are Napster and YouTube, mailing lists, Twitter and Yahoo! Answers.

Crowdsourcing systems for networking allow users to collaboratively build social network graphs, such as in the case of Facebook and LinkedIn.

Systems that allow building artifacts offer users the possibility to create textual knowledge bases, structured knowledge bases, software, and systems. Some of the main systems for building artifacts are Linux, Apache, Wikipedia, Open Mind Common Sense (OMCS), Wikipedia infoboxes/DBpedia, and Second Life.

Finally, systems that execute tasks allow, for example, content creation, cooperative debugging, searching for missing people, and elections, such as Demand Media and Associated Content.

Table 1 summarizes the surveyed crowdsourcing systems highlighting the main functionalities and the categories they belong to (if focused on evaluation, sharing, networking, building artifacts, or executing tasks).

**Table 1.** Examples of crowdsourcing systems

| Name | Description | Functionalities | Prevalent category |
|---|---|---|---|
| Amazon | platform for cloud computing and web services offered commercially | To allow users evaluating products and users using textual comments, numeric scores, or tags. | Crowdsourcing systems for evaluation |
| del.ici.ous.com | social bookmarking site for storing, searching and sharing of bookmarks | To allow users evaluating knowledge using textual comments, numeric scores, or tags. | Crowdsourcing systems for evaluation |
| Google Co-op | Google Co-op represents Google's efforts to embrace social web and social search concepts to help improve Google's search results | To allow users evaluating web pages using textual comments, numeric scores, or tags. | Crowdsourcing systems for evaluation |
| Napster | peer-to-peer file sharing Internet service that emphasized sharing audio files, typically music, encoded in MP3 format | To offer users the functionality to share audio files | Crowdsourcing systems for sharing |
| YouTube | video-sharing website | To offer users the functionality to share videos | Crowdsourcing systems for sharing |
| mailing lists | collection of names and addresses used by an individual or an organization to send material to multiple recipients | To offer users the functionality to share textual knowledge | Crowdsourcing systems for sharing |
| Twitter | online social networking service and microblogging service | To offer users the functionality to share structured knowledge and textual knowledge | Crowdsourcing systems for sharing |
| Yahoo! Answers | community-driven question-and-answer (Q&A) site | To offer users the functionality to share textual knowledge | Crowdsourcing systems for sharing |
| Facebook | social networking service | To allow users to collaboratively build social network graphs | Crowdsourcing systems for networking |
| LinkedIn | social networking website for people in professional occupations | To allow users to collaboratively build professional network graphs | Crowdsourcing systems for networking |
| Linux | computer operating system assembled under the model of free and open source software development and distribution | To offer users the possibility to create systems | Crowdsourcing for building artifacts |
| Apache | web server software | To offer users the possibility to create software | Crowdsourcing for building artifacts |
| Wikipedia | collaboratively edited, multilingual, free Internet encyclopedia | To offer users the possibility to create textual knowledge bases, structured knowledge bases | Crowdsourcing for building artifacts |
| Open Mind Common Sense (OMCS) | artificial intelligence project for building and utilizing a large commonsense knowledge base from the contributions of many thousands of people across the Web. | To allow building artifacts offer users the possibility to create textual knowledge bases, structured knowledge bases | Crowdsourcing for building artifacts |
| Wikipedia infoboxes/DBpedia | a project aiming to extract structured content from the information created as part of the Wikipedia project | To allow building artifacts offer users the possibility to create structured knowledge bases | Crowdsourcing for building artifacts |
| Second Life | online virtual world | To allow building artifacts offer users the possibility to create virtual object | Crowdsourcing for building artifacts |
| Demand Media | social-media platforms to create online content and to existing large company websites and distributes content bundled with social-media tools to outlets around the web | To allow content creation, cooperative debugging, searching for missing people, and elections | Crowdsourcing for executing tasks |
| Associated Content | a division of Yahoo that focuses on online publishing | To allow content creation, cooperative debugging, searching for missing people, and elections | Crowdsourcing for executing tasks |

Considering the context of enterprises, crowdsourcing is defined by Howe [8] as "the act of a company or institution taking a function once performed by employees and outsourcing it to an undefined (and generally large) network of people in the form of an open call."

Crowdsourcing systems allow performing a revolution in knowledge management because they give power to a distributed network of people, which is able to produce knowledge that could not be produced by a restricted number of people within the organization. By using the crowd's power, it is possible to grow services, tools and knowledge as well as to rapidly boost innovation by considering at wider number of ideas instead of someone. Advantages of crowdsourced knowledge management systems include a better use of resources across enterprises, public and private organizations, and institutions that can reduce redundant efforts and costs. Specifically, crowdsourcing can improve knowledge creation and acquisition as a greater number of more specialized people contribute to small specific tasks. In fact, a promising use of crowdsourcing is in providing information, services and tools for innovation and creativity to knowledge economy enterprises. Peng and Zhang [9] investigated the effects of online crowdsourcing on business context by underlining that crowdsourcing increases company's innovation capability and it accumulates social capital. Moreover, through the use of social technologies, more people can access to more dynamic knowledge in real-time, improving knowledge transfer and sharing.

The challenge of the proposed KRC is the engagement of the "crowd" in the co-creation and assimilation of knowledge in a way that permits its reuse and its further development for innovative and creative enterprises.

As stated in [10] "crowdsourcing involves the *management* of a community via Web-based collaborative technologies to elicit the community's knowledge and/or skill sets and thus fulfil a pre-identified business goal". Therefore, the important role of knowledge management clearly appears in crowdsourcing.

The next section provides an analysis of knowledge management systems through the perspective of collaborative knowledge.

## 3. Knowledge Management

The quick evolution of markets, connected to globalization and to the latest financial crisis, is deeply influencing the survival and competitiveness of many enterprises that need to develop their self-adapting ability to the different situations. Enterprises need to be more and more active and creative actors of the knowledge economy and also need to improve their innovation ability. Innovation is related to knowledge management and sharing, as they enable enterprises to become more and more aware of the real situation and of the available resources, and to excel taking competitive advantages [11]. As stated in [12]: "Knowledge management is not about managing technology alone, but is about managing how humans can share their knowledge effectively, using technical tools where appropriate". That is the perspective of sociotechnical systems needed to discuss Knowledge Management Systems from the point of view of human influence.

According to its source, knowledge in enterprises can be classified in internal and external knowledge as proposed by Bierly and Chakrabarthi [13]. Internal knowledge resides within the enterprise databases, within the employees of the organization, embedded in behaviours, and tends to be unique, specific and tacitly held. External knowledge comes from outside the company (e.g. from the Web, from other enterprises, from sellers, etc.) and it is widely available also to competitors. In line with that, knowledge management systems (KMSs) used by enterprises can be divided in two different categories: those that allow managing internal knowledge among individuals of the enterprises, and those that allow managing external knowledge among different enterprises, their supply chain members and customers. In their survey on KMSs in SMEs, Evangelista et al. [14] found out that these enterprises traditionally adopt internal KMSs that use relatively simple ICT tools, as (in their analysis) the largest part of knowledge management is mainly based on personal relationships and interactions within the working teams; only 23% of the SMEs, which Evangelista et al. [14] analysed, expressed their need for managing and sharing internal and external knowledge by KMSs, enabling collaborative relationships with other firms to develop common projects. This task is often supported by social networks functionalities. Caschera et al. [15] analysed methods and tools offered to community members for the knowledge management and how they support users during the exchange of knowledge and learning phases. It is important to note that the criticism of the first approach is that knowledge is mainly a capital belonging to the staff; any change and turnover of employees can produce deep changes in the company's know-how (http://www.trefle.u-bordeaux1.fr/personnel/pages/scaravetti/publis/design_2010.pdf).

A further classification of KMSs is based on the knowledge lifecycle phases. Indeed, knowledge has a dynamic nature, as it evolves over time according to a lifecycle which consists of various phases, going from three phases, as proposed by the 3-stage model of Davenport and Prusak [16] (generate, codify/ coordinate, transfer), to eight phases, as proposed by the most recent 8-stage model of King et al. [17] (Creation, Acquisition, Refinement, Memory, Transfer, Sharing, Utilization and Organizational Performance). In the 3-stage model of Davenport and Prusak, the knowledge lifecycle consists in the generation, the codification, and the transfer of knowledge. Knowledge generation refers to activities that increase the stock of organizational knowledge, which may be acquisition, dedicating resources, fusion, adaptation, and building knowledge networks. Regarding the codification, the authors state that managers must define the organisational goals to guide codification, the adequate knowledge for those goals, and the proper support for its distribution. Finally, knowledge transfer refers to activities that create market spaces and places where the trading and sharing of knowledge can happen.

The 8-stage model is shown in Figure 1.

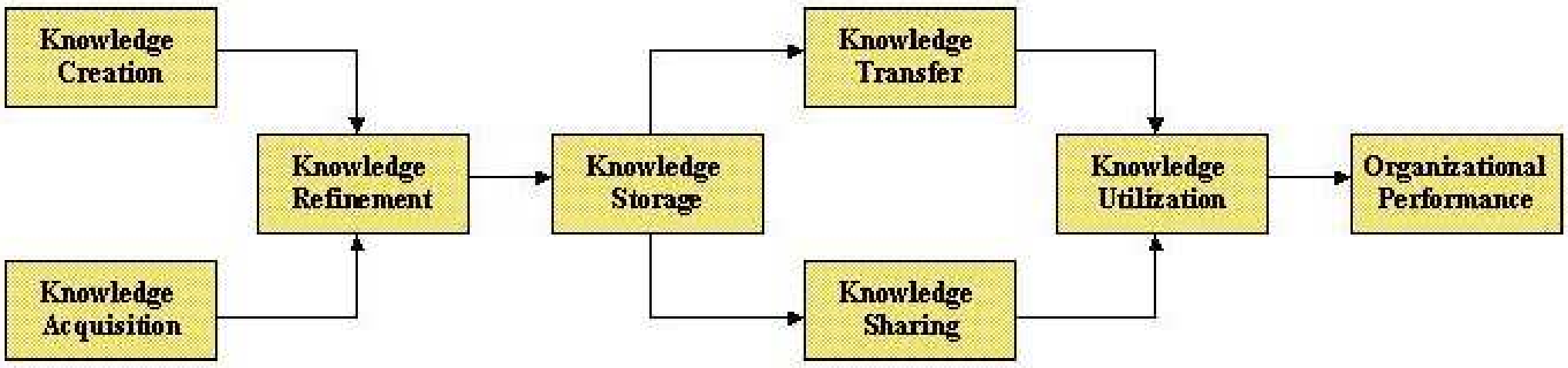

**Figure 1.** The 8-stage model of knowledge lifecycle proposed by King et al. [17]

The evolution from the model of Davenport and Prusak to the model of King et al. does not only imply that some phases have been added, but this evolution contains, implicitly and explicitly in the different phases, the concept of sharing knowledge, where the knowledge can be considered as a resource. A brief introduction to the eight knowledge lifecycle phases, along with some existing enterprise platforms and frameworks, is provided in this section, relying also on the concepts of internal and external knowledge and KMSs, which have been introduced in the previous classification.

- *Knowledge creation* involves the development of new knowledge or else the replacement of existing knowledge by new ones [18]. It is crucial to stimulate innovation in enterprises. Knowledge is usually created within the organization or in conjunction with external partners. King [17] describes creation as an activity that can consist in "socialization (the conversion of tacit knowledge to new tacit knowledge through social interactions and shared experiences), combination (creating new explicit knowledge by merging, categorizing, and synthesizing existing explicit knowledge), externalization (converting tacit knowledge to new explicit knowledge) and internalization (the creation of new tacit knowledge from explicit knowledge)." Knowledge creation and information spreading are fundamental needs to employees, investors, stakeholders, as well as any other person who belongs to the enterprise. An example of internal knowledge creation process among people in the enterprise is the creation of software design patterns. A framework for software design patterns is provided in [19]. This framework allows simultaneously activating multiple views [20] according to the demand of each user at any time. Each view simultaneously and independently presents the same information from the model with less development costs and less controller workload.
- *Knowledge acquisition* refers to the search, the recognition and the assimilation of potentially valuable knowledge, often from outside the organization and the Web [21]. Considering the knowledge acquisition process, as knowledge is nowadays constantly updated, enterprises that want to maintain their competitiveness must be able to continuously acquire new knowledge, both internal and external, and skills. PIKR (Production and Innovation Knowledge Repository) [22] is an example of a platform which functionalities mainly support the knowledge acquisition process and boost open innovation (this concept is shaping up crowdsourcing approaches, which has been detailed in the previous section), the production and the development activities of virtual enterprises. Virtual enterprises are, indeed, the target users of this platform, which benefit from a set of ontologies in order to semantically describe the knowledge resources of an enterprise, and semantics-based services for accessing and reasoning over such descriptions. Specifically, the PIKR provides the following functionalities for external knowledge management to users: i) a wiki-like environment to define semantic descriptors of knowledge resources (documental and expertise resources); ii) navigating resources and retrieving information through semantic keyword-based search and query services; iii) reasoning over the knowledge resources using inference services.

- *Knowledge refinement* involves processes and mechanisms that are used to select, filter, purify and optimize knowledge for inclusion in various storage media [17]. Knowledge, as it is created, should not be so useful for members within the organization, therefore a set of processes, which includes explication, evaluation, selection, and codification, must be performed in order to prepare the newly created knowledge to get into the organisation memory. One of the systems, which mainly addresses this phase, is IMPA (Innovation Management PlAtform) [23]; it is a web-based collaborative workflow-based innovation management application that is mainly focused on the refinement phase of the knowledge lifecycle. The IMPA has been designed to help enterprises target users of this platform, fostering break-through innovation. The main functionalities of IMPA are to: i) capture, edit, share and develop new ideas in a collaborative environment; ii) group knowledge (ideas, problems, solutions, and information) by categories and keywords to find potential synergies and match problems to solutions; iii) use feedback loops to periodically re-evaluate ideas/proposed and realized projects based on new information; iv) provide review and rebuttal/idea owner's response process stages; v) allow cross-functional/cross-departmental sharing, discussion and development. Functionalities ii-iv are mainly devoted to knowledge refinement, as they allow the categorization, the evaluation, the selection, and the revision of new ideas. This platform applies quantifiable measures (key performance indicators) to compare the innovation performance of the organizations. This application is able to manage both internal and external knowledge.
- *Knowledge storage* involves the acquisition of knowledge from organizational members and/or external sources, the coding and the indexing of knowledge, and the capture of it [24]. Knowledge storage is one of the fundamental activities of KMSs. Nowadays, almost all SMEs search for external information, in different sources such as the web, the patent databases, etc. More than 90% of SMEs uses Internet search as main source of knowledge [25]. Retrieved information needs to be structured, collected, and managed in order to be re-used by all members within the enterprise. To achieve this aim, several platforms have been developed in literature over the last few years. INSEARCH [26] is a platform for Enterprise Semantic Search that integrates analytical natural language analysis tools with collaborative and semantic management systems for internal and external knowledge. The target users of this platform are the members of networked enterprises that produce, reuse and validate their knowledge in a shared manner. The functionalities offered by the platform allow organizing knowledge in a semantic way so that the contextual search through the system is supported. Specifically, the basic search features from web sources, scientific papers and patents are extended with a navigation in linked search results, recommendations, semantic bookmarking, intelligent document analysis, and website monitoring.
- *Knowledge transfer* involves the communication of knowledge from a sender to a known receiver [27].
- *Knowledge sharing* is a less-focused dissemination, such as through a repository, to people who are often unknown to the contributor [28]. Knowledge sharing may arise as both internal and external knowledge. The use of knowledge sharing mechanisms enables enterprises to realize online customer services, leading to higher customer service efficiency. Knowledge sharing can be performed by the sharing between different departments in enterprises, or by the sharing between enterprises and their supply chain members; or else by the knowledge sharing between enterprises and customers [29]. In [30] a framework that combines mechanism of knowledge sharing and network technology is provided. This online service system provides functionalities that promote product improvement and customer service. This framework allows enterprises to transfer knowledge about product and service to consumers through virtual community, and on the other hand, it allows consumers to share their knowledge during the use with enterprises. This framework combines the functionalities of the knowledge management with the functionalities of the online service provider. The knowledge services include: knowledge library; knowledge verification and evaluation; knowledge sharing; knowledge retrieval, real-time interaction, and knowledge map. The advantages of this kind of framework are that it brings convenience to customers for the use of product and it can reduce the pressure of enterprise customer service work. Yiqing et al. [31] investigate the use of semantic web and ontology technologies for an effective information exchange and knowledge sharing in supply chains.
- *Knowledge utilization* involves the use through elaboration, infusion, and thoroughness. The re-use of knowledge may be implied in several contexts, e.g. to facilitate innovation, collective learning, individual learning, and/or collaborative problem solving [32]. Knowledge can be obtained by different ways, such as conventional class participation, reading, or advanced network technologies. E-learning is an efficient and effective learning approach that is able to provide the knowledge and skill in the shortest period of time. E-learning are capacity building tools that can effectively help SMEs to improve their employees' work performance and to lower staff training costs by increasing in turnover and by reducing in manpower [33]. In [34] an e-learning service provider framework is described underlining how it incorporates learning content provider, network infrastructure provider, and learning management system provider. This framework aims at enhancing teaching quality of e-learning systems and at improving teaching content interchanges and management. This purpose is achieved by functionalities that allow updating, storing, retrieving, distributing, and sharing any time teaching materials and information, and transmitting different media types including texts, pictures, sounds, and images. Moreover, it provides network communication mechanisms. Although the traditional process of knowledge management does have measurable benefits (e.g. improved competency, efficiency, decision making, learning, innovation, and increase in revenue, as investigated in [35], it is cumbersome, expensive and slow, because considerable time and efforts are required to allow enterprises articulating their knowledge.

- *Organizational performance* is at the end of the knowledge lifecycle and consists in all the activities that allow to evaluate the impact of KM initiatives in the organizational performance.

Table 2 summarizes the surveyed enterprise's knowledge management systems highlighting the main functionalities and the categories they belong to (internal or external KMSs, and focused on knowledge creation, acquisition, refinement, storage, transfer, sharing, or utilization).

**Table 2.** Examples of knowledge management systems

| **Name** | **Description** | **Functionalities** | **Prevalent category** |
|---|---|---|---|
| INSEARCH Enterprise Semantic Search | INSEARCH integrates analytical natural language analysis tools, robust adaptive methods and semantic document management systems | -access, creation and refinement of description of domains that act as collectors for documents<br>-semantic bookmarking and annotation<br>- contextual search | Knowledge storage |
| PIKR Production and Innovation Knowledge Repository | PIKR is a knowledge-based platform that supports open innovation in virtual enterprises. | - keyword-based search<br>- query<br>- reasoning | Knowledge acquisition |
| ASP-based e-Learning service provider framework | It incorporates a learning content provider, a network infrastructure provider, and a learning management system provider. | -Teaching materials and information can be updated, stored, retrieved, distributed, or shared any time.<br>-network communication mechanisms<br>-transmission of different media types including texts, pictures, sounds, and images | Knowledge utilization |
| Model-View-Controller (MVC) architecture | Model-View-Controller architecture is widely used software design because it decouples the application object and screen presentation, and it increases flexibility and reusability through the way user interface reacts to user. | -development team to communicate with database<br>-format the heavy reports according to the demand of each and every user every time | Knowledge creation |
| A Enterprise Online Service System | Enterprise online service system allows: enterprises use to transmit knowledge about products and services to consumers through virtual community; consumers to share their knowledge during the use with enterprises. This system promotes the products improvement and customer services. | - knowledge library<br>- knowledge verification and evaluation<br>- knowledge sharing<br>- knowledge retrieval<br>- real-time interaction.<br>- knowledge map | Knowledge transfer<br><br>Knowledge sharing |
| IMPA Web-Based Innovation Management Software Platform | IMPA is a Web-2 collaborative workflow-based Innovation management application. IMPA is designed to help generate, identify, define and select innovation projects. It also supports and enhances incremental innovation. | -Capture, edit, and mature ideas collaboratively<br>- Group ideas by categories and keywords to find potential synergies and match problems to solutions<br>-Use feedback loops to periodically re-evaluate ideas/proposed and realized projects based on new information<br>- Provide review and rebuttal / idea owner's response process stages<br>-Cross-functional/cross-departmental sharing, discussion and development | Knowledge refinement |

The emergence of the Internet and Web 2.0 and 3.0 technologies significantly lowered the costs and the time of knowledge management process through user's participation. Specifically, crowdsourcing techniques, which allow outsourcing a task to the "crowd", enable people who are involved in the knowledge management process to co-produce knowledge and tools according to the competencies and motivations of each one [36]. Shifting from the 3-stages KM to the 8-stages KM some concepts that characterize crowdsourcing have been embedded in KMSs. In fact, they have been enriched by states that allow co-producing knowledge, also involving crowd.

The next section presents the KRC underlining its functionalities in the perspective of a crowdsourced knowledge management system.

## 4. KRC: the Knowing Resource Centre

Technology and knowledge are key factors of innovation, and knowledge sharing and expertise can represent a driving force for cultural heritage industries and organizations, craft firms, high-tech companies, higher education institutions, research centers and advanced producer services.

The KnowInG project deals with the common need of the MEDiterranean regions to find and experiment a better governance of innovation in the global challenges of the current economy. The activities of the project were addressed to enhance the cooperation of the key institutional and economic actors promoting the "knowledge economy" by launching a transnational dialogue, starting from the awareness that regional growth passes through the collaboration among the key actors to achieve growth objectives. Dealing with innovation tools, the KnowInG project refers to all services, actions and objects that policies and actors make available for supporting innovation. These tools are used in order to promote and to generate creativity, to break existing patterns, to stimulate the imagination and, to improve situations in which the creative ideas are generated.

KnowInG has developed KRC, starting from PLAKSS, with the awareness that creativity arises through the confluence of knowledge, creative thinking and motivation of involved actors; these elements are frequently relevant at local level in small communities, where common social beliefs and common values create the trust needed to share knowledge, information and services. KRC is based on a collective vision of resources (i.e. information, knowledge and services) involving the concept of crowd.

Companies use crowdsourcing for matching needs and competencies, involving crowd. This approach is more and more frequently adopted, as the pervasive use of Web 2.0 technologies allows following more participative strategies in design, creation and marketing processes. An example is represented by companies that involve customers in the design process of new products, or in defining new strategies. Each customer is actively involved by the company to add a resource. The idea of involving the crowd has been completely adopted by KRC, as it is a tool with which companies, as well as people interested in creative and innovative activities, can share resources.

KRC is an intelligent Web platform aiming at providing information and services to enterprises, research centres, universities, public and local administrations and people involved in creative and knowledge production processes in order to support innovation and creativity. It is a crowdsourced knowledge management system based on an adaptive and interactive participation of people that play both an active and a passive role, in providing and using information, data and services shared by the community.

KRC stimulates innovation processes helping the local and regional development and its projection at transnational level in the MED area; it acts by engaging the "crowd" in the co-creation and assimilation of resources, and by allowing actors of innovation to identify problems and needs, matching them with the existing solutions and resources. KRC acts as a multiplier of innovation as it allows aggregating, accessing, and sharing services, information and knowledge already available in pre-existing tools (e.g. on-line helpdesks, databases, wikis, communities on social networks) in an organized and unified manner, as well as creating a personal repository of services, information and knowledge, which can be shared in the community built in the KRC.

Resources, which can be created, acquired and shared, concern actors (e.g. institutions, universities, companies, research centers, incubators, scientific and technological parks, etc.), policies, funding tools and services (e.g. help desks, directories, databases, etc.) on innovation and creativity, for regions and countries in the MED area, and more generally in Europe, providing opportunities at a local, a national and a transnational level.

All information and services are organized and accessed using a set of concepts to facilitate their access and fruition. These concepts have been structured in categories and topics identified during the project activities on the basis of a questionnaire administrated to stakeholders and actors of innovation. The results of the survey are categories (e.g. Institutions, Universities, Research Centers, Companies, Transnational founding tools, Transnational policies) and topics (e.g. Innovation, Knowledge Economy, Creativity, Technology Transfer). Each user, starting from the pre-defined set of concepts, can add her/his own set of concepts to organize her/his knowledge. The target users of KRC are all possible actors of innovation such as: public and private organizations, entrepreneurs, professors, creativity and knowledge builder, common people interested in innovation. Each user is mainly a member of a community by sharing resources; she/he must be registered on the KRC and she/he can play both the role of problem and solution provider. KRC integrates some functionalities offered by KMSs with those offered by crowdsourcing systems as specified below. Specifically, KRC allows knowledge creation, acquisition, storage and refinement by:

- Browsing on-line information and services by using the most popular search engines (e.g. Google).
- Browsing shared information and services stored in the KRC and organized according to an ontology of concepts.
- Browsing activities and information in online communities (i.e. Facebook, Twitter and LinkedIn).
- Associating information and services with categories and topics organizing them. In particular, the different resources are organised in categories (e.g. Institutions, Universities, Research Centers, Companies, Transnational founding tools, Transnational policies) and topics (e.g. Innovation, Knowledge Economy, Creativity, Technology Transfer). These categories are classified according to the different topics, which can be defined by each user for giving concepts that organize knowledge.
- Databases that contain sets of data about actors (such as companies, no profit organization, local administration), projects, and their web links on innovation, creativity, research and technology development; directories that provide lists of blogs, news, web links, essays, and other sources of information and knowledge about innovation, creativity, and knowledge based economy; services such as helpdesks for accessing funding tools, and supporting SMEs in defining strategies, business plans and programs for transnational, national and local funding opportunities; social networks, which include Facebook pages and groups about KnowInG project and KnowInG national communities.

- Creation of personal databases with on-line information and services organizing the acquired knowledge in shared databases, directories, funding tools, services organizing the acquired information and services in personal databases.
- Providing a unique e-mail service that allows managing all the personal accounts in a unified tool, as well as accessing information and services of interest contained in the e-mail.
- Selecting and filtering information and services stored in the platform by users' profile information to optimize the search process for the closest information and services according to the user profile.

Moreover, KRC facilitates knowledge creation by knowledge transfer and sharing, knowledge utilization, social networking and building artifact functionalities. In fact, each user can share with the community all or part of the information and the services contained in the personal resources defined in the KRC. Each one can also re-use knowledge:

- Combining the results of searching resources stored in the platform or available online to find and elaborate new ideas.
- Re-using the profile information to reduce the search space for the closest information and services according to the user profile.

KRC implements a virtual community from social network groups by creating a community of interest by joining individuals as well as pre-existing networks on the basis of features of human behaviours, interests and user profiles.

Moreover KRC allows building artifact functionalities by:

- Building personalized and structured knowledge bases, by selecting data and information from available resources, categorizing them and integrating in new knowledge.
- Building structured knowledge bases by adding URL and supply attribute-value pairs into the KRC.

Table 3 provides a description of the KRC, its main functionalities and the categories of KMSs and crowdsourcing systems (introduced in the previous sections) they refer to.

After six months from the beginning of the KRC use, 1973 web resources are shared between 180 users; these web resources are distributed on categories and topics already described according to the two-dimensional representation of Figure 2. It shows that Companies, Universities/Research Centres, Associations and Institutions were more involved in sharing their resources, in particular considering topics on Creativity and Innovation. It can also be noted that the presence of on-line communities related to creativity and innovation is prevalent. This result is also connected with the fact that the largest part of the resources shared in the KRC were collected by actors involved in the activities of KnowInG, such as companies, Universities and Research Centres, all focused on creativity and innovation, as KnowInG chooses tourism, fashion, pottery, crafts, videogames and green energy as focuses in the regions involved in the project.

**Table 3.** The KRC's description, functionalities and categories

| Name | Description | Functionalities | Category |
|---|---|---|---|
| THE KNOWING RESOURCE CENTRE | It can be conceived as a hub of information, data, knowledge and services for the community of interest. It is a platform that stimulates and manages Sustainable Social Innovation by combining open | - browsing on-line information and services by using a tool that is based on the most popular search engine (Google,..)<br>- browsing shared information and services stored in the platform and organized according to an ontology of concepts<br>- browsing activities and information in the online communities (facebook, twitter and LinkedIn) | knowledge acquisition |
| | | - enabling people to profile themselves, on the base of pre-defined profiles<br>- defining a new profile providing its features<br>- providing the profile information | knowledge creation |
| | online social media, services and the creation and management of distributed knowledge and data. | - enabling people to organizing the retrieved on-line knowledge classifying it in: databases, directories, funding tools, services, social networks<br>- enabling people to organize the retrieved on-line information and services in personal databases<br>- allowing users to have a unique e-mail service that allows to manage all the personal accounts in a unified tool, accessing also information and services of interest contained in the e-mail<br>- definition of databases with set of data about projects and web links on innovation, creativity, research and technology development<br>- definition of personal databases with on-line information and services | Knowledge storage |
| | | - selecting and filtering information and services stored in the platform using the profile information | knowledge refinement |
| | | - enabling people to share with the community all or part of information, services and tools contained in the personal databases<br>- enabling people to share knowledge according to different topics, available on Facebook, Twitter and LinkedIn | knowledge sharing |
| | | -combining the results of searching resources stored in the platform or available online to find and elaborate new ideas<br>-re-using the profile information to reduce the search space | Knowledge utilization |
| | | -creating a community of interest joining individuals as well as pre- existing networks on the base of features of human behaviours, interests and user profiles | Networking |
| | | - building of personalized structured knowledge bases selecting data and information from available resources, categorizing them and integrating in new knowledge<br>- building a structured knowledge bases by adding URL and supply attribute-value pairs into the KRC | Building artifact |

The wide use of KRC, its strength and weakness are mainly connected with the involvement of the crowd, by enlarging the use to build communities that share different services. The usefulness and strength of KRC consists in its capability of connecting existing resources in a single tool by functioning as a hub of information, knowledge and services. Each user can use these resources and, at the same time, can enhance these resources by sharing their own knowledge.

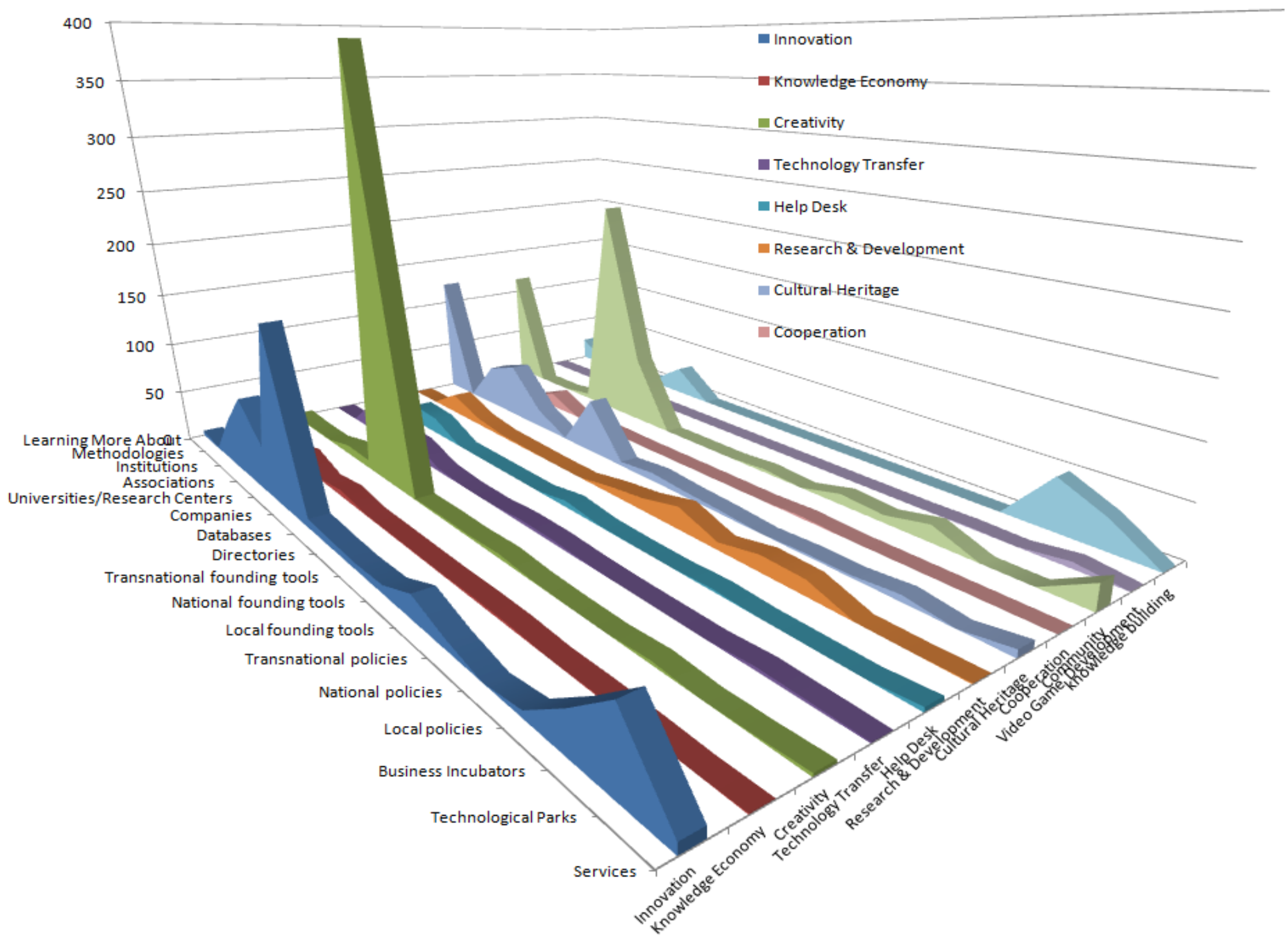

**Figure 2.** Distribution of the KRC web resources per category and topic after six months from the beginning of the use of KRC.

## 5. Conclusion

In the next future, we can expect that knowledge management as well as knowledge and services sharing will become even more collaborative, by giving access to more dynamic contents, being sometimes co-created and developed by the "crowd". This will foster the creativity and the innovation process; and the competitiveness of enterprises. Therefore, a promising use of crowdsourcing is in the area of management of information, services and tools for knowledge economy enterprises in order to promote their innovation and creativity.

KRC, which has developed throughout the project "Knowledge Intelligence and Innovation for a sustainable Growth - KnowInG", has been conceived aiming at engaging companies, institutions, organizations and the "crowd" in the co-creation and assimilation of knowledge by sharing resources (i.e. information, knowledge and services) in a way that permits its re-use and its further development for innovative and creative enterprises. KRC is a crowdsourcing platform, i.e. an open innovation tool allowing people, companies, research institutions, universities and organizations to improve and increase knowledge sharing, creativity and innovation beyond the project.

## 6. References

[1] Estellés-Arolas, E., González-Ladrón-de-Guevara, F. "Towards an Integrated Crowdsourcing Definition", Journal of Information Science, SAGE, vol. 38, no.2, pp. 189–200, 2013.

[2] Zhao, Y., Qinghua Z. "Evaluation on Crowdsourcing Research: Current Status and Future Direction", Information Systems Frontiers, Springer US, Online First, published online April 11, 2012. http://link.springer.com/article/10.1007%2Fs10796-012-9350-4. (Accessed October 11, 2013).

[3] Chesbrough, H., West, J., Vanhaverbeke, W. "Open Innovation: Researching a New Paradigm", Oxford University Press, 2006.
[4] Seltzer, E., Mahmoudi, D. "Citizen Participation, Open Innovation, and Crowdsourcing: Challenges and Opportunities for Planning", Journal of Planning Literature. SAGE, Online First Version of Record - Dec 10, 2012.. http://jpl.sagepub.com/content/early/2012/12/10/0885412212469112.full.pdf+html (Accessed April 12, 2013)
[5] Dearden, A., Rizvi, H. "A deeply embedded sociotechnical strategy for designing ICT for development", Journal of SocioTechnology and Knowledge Development, vol. 1, no. 4, pp. 52 - 70, 2009.
[6] Coakes, J., Coakes, E. "Sociotechnical Concepts applied to Information Systems", Management Information Systems, Vol. 7 of the Blackwell Encyclopedia of Management, pp. 281-286, 2005.
[7] Doan, A., Ramakrishnan, R., Halevy. A. "Crowdsourcing systems on the World-Wide Web". Communications of the ACM, vol. 54, no. 4, pp.86–96, 2011.
[8] Howe, J. "Crowdsourcing: A Definition", Crowdsourcing Blog, 2006. [Online] http://crowdsourcing.typepad.com/cs/2006/06/crowdsourcing_a.html (Accessed February 15, 2013).
[9] Ling PENG, "Social Capital in Online Crowdsourcing Participation: An Empirical Study", AISS, Vol. 5, No. 4, pp. 222-229, 2013.
[10] Saxton, G. D., Oh, O., Kishore, R. "Rules of Crowdsourcing: Models, Issues, and Systems of Control". Information Systems Management, 2011.
[11] Huang M., Sun, B. "Research on modeling and implementation of knowledge management system in virtue enterprise", Proceedings of the Eighth International Conference on Machine Learning and Cybernetics, 2009.
[12] Lehaney, B., Clarke, S., Coakes, E., Jack, G. "Sociotechnical Systems and Knowledge Management", Beyond Knowledge Management, PA: Idea Group Publishing, Hershey, pp. 31-75, 2004.
[13] Bierly, P., Chakrabarthi, A. "Exploration and exploitation in organizational learning", Organizational Science, vol. 2, no. 1, pp. 71-87, 1996.
[14] Evangelista, P., Esposito, E., Lauro, V., Raffa, M. "The Adoption of Knowledge Management Systems in Small Firms", Electronic Journal of Knowledge Management, vol. 8, no. 1, pp. 33 – 42, 2010.
[15] Caschera M.C., D'Ulizia A., Ferri F., Grifoni P. "Knowledge Management and Interaction in Virtual Communities", Semantic Knowledge Management: an Ontology-based Framework, IGI Publishing, (Zilli, Damiani, Ceravolo, Corallo and Elia editors), pp. 216-232, 2009.
[16] Davenport, T.H., Prusak, L. "Working knowledge: How organizations manage what they know", Harvard Business School Pr., 1998.
[17] King, W.R. "Knowledge management and organizational learning", Knowledge Management and Organizational Learning, Springer US, pp. 3-13, 2009.
[18] Nonaka, I. "A dynamic theory of organizational knowledge creation". Organizational Science, vol. 5, no. 1, pp. 14–37, 1994.
[19] Ahmad, F., Najam, A., Ahmed, Z. "MVC Architectural Pattern for Enterprise Information System's Reporting Subsystem", J2EE International Conference on Future Trends in Computing & Communication Technologies, pp. 122-124, 2012.
[20] Kotek. B. "MVC design pattern brings about better organization and code reuse", 2002 [Online]. http://www.techrepublic.com/article/mvc-design-pattern-brings-about-better-organization-and-code-reuse/1049862 (Accessed February 15, 2013)
[21] Huber, G.P. "Organizational learning: The contributing processes and the literatures", Organization Science, vol. 2, no. 1, pp. 88–115, 1991.
[22] Taglino, F., Smith, F., Proietti, M. "Knowledge-Based Support to Business Innovation", Proceedings of NGEBIS Workshop, pp. 37-44. 2012.
[23] Altran HG, "IMPA: Web-Based Innovation Management Software Platform", technical Report, 2012, [Online] http://www.altran.ch/fileadmin/medias/CH.altran.ch/files/FP7_A4_web_2.pdf (Accessed February 15, 2013)

[24] Alavi, M., Leidner D.E. "Review: Knowledge Management and Knowledge Management Systems: Conceptual Foundations and Research Issues". MIS Quarterly, vol 25, no 1, pp.107-136, 2001.
[25] Cocchi, L., Bohm, K. "Deliverable 2.2: Analysis of functional and market information", TECH-IT-EASY, 2009.
[26] De Cao, D., Storch V., Croce, D., Basili, R. "INSEARCH: A platform for. Enterprise Semantic Search". Proceedings of IIR Workshop, pp. 104-115, 2013.
[27] King, W.R. "Knowledge transfer", Encyclopedia of knowledge management, pp. 538–543. Hershey, PA: Idea Group Publishing, 2006.
[28] King, W.R. "Knowledge sharing", Encyclopedia of knowledge management, pp. 493–498. Hershey, PA: Idea Group Publishing, 2006.
[29] Jiewang, C., Qianping, F. "Comment on the Theory and Practice of Knowledge Sharing in China and Abroad". Information Studies: Theory & Application, Vol. 5, No. 30, pp. 705-709, 2007.
[30] Jianlei, B., Lin Z. "A Enterprise Online Service System: Knowledge Sharing Mechanism Based on Virtual Community", Proceedings of the 2010 International Symposium —Technical Innovation of Industrial Transformation and Structural Adjustment, pp. 43-49, 2010.
[31] Yiqing Lu, Lu Liu, Chen Li, "Knowledge Sharing and Integration in Supply Chain Based on Semantic Web", AISS: Advances in Information Sciences and Service Sciences, vol. 4, no. 22, pp. 524 -532, 2012.
[32] King, W.R. "Communications and information processing as a critical success factor in the effective knowledge organization". International Journal of Business Information Systems, vol. 10, no.5, pp. 31–52, 2005.
[33] Boursa, C., Hornig, G., Triantafillou, V., Taiatsos, T. "Architectures Supporting E-Learning Through Collaborative Virtual Environments: The Case of INVITE". Proceedings of IEEE International Conference on Advanced Learning Technologies, pp. 13-16, 2001.
[34] Chang,C. and Cheng, C. "ASP-Based E-Learning Services for Business Education and Training", The Journal of Human Resource and Adult Learning, vol. 3, no. 2, pp. 180-188, 2007.
[35] Rasheed, N. "The impact of knowledge management on SMEs", Knowledge Board, pp 1-15, 2005.
[36] Howe, J. "Crowdsourcing: Why the Power of the Crowd is Driving the Future of Business", 2009. [Online]. http://crowdsourcing.typepad.com/ (Accessed February 15, 2013).